\documentclass[12pt,aps,nofootinbib]{article} %,draft
\usepackage{amsmath,amssymb,physymb}
\usepackage[mathscr]{euscript}
 \usepackage[pdftex]{graphicx}\usepackage{epstopdf}
\newcommand*{\bi}[1]{\boldsymbol{#1}}
\newcommand*{\btau}{\bi{\tau}}

\DeclareMathAlphabet{\mathpzc}{OT1}{pzc}{mb}{it}
\newcounter{aaa}

\newenvironment{teor*}[2][{}]{\begin{trivlist}%
\labelsep=0pt\item[\bfseries #2. ]#1}
{\end{trivlist}}

\newcommand{\ssy}[5]{#1,    #2 {\bf #3}, #5 (#4)\rlap{.}}

\newenvironment{proof}[1][.
]{\par\noindent\emph{Proof#1}}{\par\nopagebreak
\hfill$\square$\par}

\newcommand{\rmd}{\mathrm{d}}

\newcommand{\varg}{g}

\newcommand*{\comp}{{\mathcal K}}

\newcommand{\ingh}{M^\mathrm{in}}
\title{Yet another proof of Hawking and Ellis's Lemma 8.5.5}

\author{S. Krasnikov\thanks{Email: S.V.Krasnikov@mail.ru}}%
\date{}
\begin{document}
\maketitle
\begin{abstract}
The fact that the null generators of a future Cauchy horizon are past complete was first proved by Hawking and Ellis \cite{HawEl}. Then Budzy\'n\-ski, Kondracki, and  Kr\'olak outlined a proof free from the error found in the original one \cite{BKK}. Finally,   Minguzzi has recently published his version of the proof \cite{Ming} patching a previously unnoticed hole in the preceding two. I am not aware of any flaws in that last proof, but it is  quite difficult. In this note I present a simpler one.
 \end{abstract}
%\pacs{04.20.Gz}
\maketitle
\section{Introduction}
Let $H^+$ denote a future Cauchy horizon. A lemma by Hawking and Ellis says
\begin{teor*}{Lemma 8.5.5 of \cite{HawEl}}
    If $H^+(\mathcal Q)$ is compact for a partial Cauchy surface $\mathcal Q$, then the null
geodesic generating segments of $H^+(\mathcal Q)$ are geodesically complete in
the past direction.
\end{teor*}
The lemma itself has never been doubted (to my knowledge), but the  proof offered in \cite{HawEl} was found to be flawed, see \cite{Ming} and  references therein. To improve this situation --- which is important, because the lemma is a popular tool in mathematical relativity --- Minguzzi has recently published a new, more accurate, proof of the lemma (or, to be precise, of some strengthening of it).

In this comment I present yet another proof of the same  fact. The reason for doing this is that my version is much simpler (partly because its major part  is replaced by a reference to a lemma proved elsewhere).

\section{The proposition and its proof}

In a spacetime $M$ consider a  past inextendible null curve $\gamma$   totally  imprisoned in a compact set $\comp$. Pick   a smooth unit timelike future directed vector field $ \btau $ on $M$  and define (uniquely up to an additive constant) the ``arc length parameter''
$ l $ on $\gamma$ by the requirement
\begin{equation}\label{eq:sphera}
\varg(\partial_ l,\btau)=-1.
\end{equation}
 In addition to  $ l $
define on $\gamma$  an \emph{affine} parameter $ s$  so that $\partial_ s$ is future directed and $s=0$ at $l=0$.
Now $\gamma$  is characterized by the (evidently negative) function
\[
h\equiv \varg(\partial_s,\btau),
\]
which relates $ l $ to $ s$:
\begin{equation}\label{def:h}
h=-\frac{\rmd  l }{\rmd s},\qquad  s( l )=\int_ l
^0\frac{\rmd  \breve l }{h(\breve l )}.
\end{equation}
As is proven in \cite{cgch}
  \begin{equation}\label{eq: h'/h bound}
  h'/h\quad\mbox{is bounded on } \gamma.
\end{equation}
Further, a few minor changes --- a past inextendible $\gamma (l)$ with $l \in (-\infty,0]$  instead of the future inextendible $\gamma(l)$ with $l \in [0,\infty)$  and an arbitrary compact $\comp$ instead of some specific $\mathcal L$ --- leave \cite[Lemma~8]{cgch} valid while bringing it to the following form.

\begin{teor*}{Lemma 8 of \cite{cgch}} \label{lem: part of 8.5.5}
Assume  $h(l)$ is such that there exists a   smooth  function $f(l)$ defined at non-positive $l$ and obeying for some non-negative constants   $c_1$, $\underline f $, $\overline f $ the inequalities
\begin{equation*}
 \underline f \leq f
\leq  \overline f       ,\qquad      |f'/f|<\infty     \nonumber
\end{equation*}
and
\begin{equation}\label{eq:ogr na f}
h'/h< - f'/f -c_1  f
 ,\qquad
\forall   l \leq 0.
\end{equation}
Then there is a timelike   past inextendible  curve $\gamma_{\varkappa_0}$ which is obtained by moving each point of $\gamma$ to the past along the integral curves of $\btau$ and which is  totally  imprisoned in a compact set $\mathcal O$.
\end{teor*}

\begin{teor*}{Proposition}\label{prop:8.5.5} If $\comp$ is a subset of  the boundary of
a globally hyperbolic past set $\ingh$,
then  $\gamma$ is past complete.
\end{teor*}
\begin{proof} Suppose the lemma is false and $\gamma$ is past incomplete. This would mean that the affine parameter $ s$ is bounded from below and, correspondingly,
the integral  \eqref{def:h} converges at $ l \to-\infty$.
Which allows one to define the following smooth positive function on  $(-\infty,0]$
\begin{equation}\label{eq:def x}
f( l )\equiv \frac1h\left[-\int_ l ^0\frac{\rmd  \breve l }{h(\breve l )} + 2
\int_{-\infty}^0\frac{\rmd  \breve l }{h( \breve l )}\right]^{-1}.
\end{equation}
$f$ so defined satisfies the equation
\begin{equation}\label{eq: ln(fh)'}
f'/f +h'/h=-f
\end{equation}
and consequently, condition~\eqref{eq:ogr na f} holds.

As $h$ is negative, the boundedness of the integral  \eqref{def:h} provides a simple estimate
\[ %\Bigl(\frac{1 }{h(  l )}\Bigr)^2
\infty >-\int_ {-\infty}
^0\frac{\rmd  \breve l }{h(\breve l )} > - \frac{1/h(l)}{(|h'/h|)^{\mathrm{max}}},
\qquad \forall l \in (-\infty,0],
\]
which implies, due to \eqref{eq: h'/h bound}, that $1/h$ is bounded. It follows then from \eqref{eq:def x} that $f$  is bounded too. Finally, the just proven boundedness of $f$ combined with \eqref{eq: ln(fh)'} and \eqref{eq: h'/h bound} implies the boundedness of $f'/f$.
Thus all the conditions of Lemma~\ref{lem: part of 8.5.5}  are fulfilled and the corresponding variation transforms $\gamma$ into
a past inextendible timelike curve $\gamma_{\varkappa_0}$. The latter being timelike lies entirely in the closed (due to the globally hyperbolicity of $\ingh$, to which $\gamma_{\varkappa_0}(0)$ belongs) set $J^-(\gamma_{\varkappa_0}(0))\subset \ingh$ (the inclusion follows from the fact that $\ingh$ is a past set). Thus, $\gamma_{\varkappa_0}$ is   totally imprisoned in the compact  subset $\mathcal O\cap J^-(\gamma_{\varkappa_0}(0))$ of the globally hyperbolic spacetime $\ingh$, which is forbidden by \cite[proposition 6.4.7]{HawEl}.
\end{proof}

\end{document}